\documentclass[11pt]{article}

\usepackage{hyperref}
\usepackage{AAOAJ-2022-LPFSKE}
\usepackage{mathptmx}
\usepackage{authblk}
\usepackage{lipsum}
\usepackage{xurl}

\newcommand{\mysize}[2]{\fontsize{#1}{#2}\selectfont}

\makeatletter
\renewcommand{\maketitle}{\bgroup\setlength{\parindent}{0pt}
	\begin{flushleft}
		\textbf{\@title}
		
		\@author
	\end{flushleft}\egroup
}
\makeatother

\title{\mysize{15pt}{17pt}Symbolic Knowledge Extraction from Opaque Predictors Applied to Cosmic-Ray Data Gathered with LISA Pathfinder
}

\author[]{\mysize{11pt}{13pt}Federico Sabbatini \hspace{2em} Catia Grimani}

\affil[]{\mysize{10pt}{12pt}Department of Pure and Applied Sciences, University of Urbino Carlo Bo, Italy\\INFN, Section in Florence,  Italy}

\affil[]{\mysize{10pt}{12pt}Federico Sabbatini (corresponding author), University of Urbino, Via S. Chiara, 27, 61029, Urbino, Italy, \href{mailto:f.sabbatini1@campus.uniurb.it}{f.sabbatini1@campus.uniurb.it}\\Catia Grimani, University of Urbino, Via S. Chiara, 27, 61029, Urbino, Italy, \href{mailto:catia.grimani@uniurb.it}{catia.grimani@uniurb.it}}

\affil{\mysize{11pt}{13pt} Research Article}

\usepackage{fancyhdr}
\begin{document}
	
\maketitle

%\begin{abstract}

{
\mysize{10pt}{12pt}

\section*{Abstract}

Machine learning models are nowadays ubiquitous in space missions, performing a wide variety of tasks ranging from the prediction of multivariate time series through the detection of specific patterns in the input data.
Adopted models are usually deep neural networks or other complex machine learning algorithms providing predictions that are opaque, i.e., human users are not allowed to understand the rationale behind the provided predictions.
Several techniques exist in the literature to combine the impressive predictive performance of opaque machine learning models with human-intelligible prediction explanations, as for instance the application of symbolic knowledge extraction procedures.

In this paper are reported the results of different knowledge extractors applied to an ensemble predictor capable of reproducing cosmic-ray data gathered on board the LISA Pathfinder space mission.
A discussion about the readability/fidelity trade-off of the extracted knowledge is also presented.

\textbf{Keywords:} LISA Pathfinder, ensemble regressor, explainable AI, symbolic knowledge extraction, PSyKE

\textbf{Abbreviations:} Galactic cosmic ray (GCR), LISA Pathfinder (LPF), Symbolic knowledge extraction (SKE)

%\end{abstract}

%\section*{Abbreviations}

%GCR --- Galactic cosmic ray

%\noindent LPF --- LISA Pathfinder

%\noindent SKE --- Symbolic knowledge extraction

\section{Introduction}

Data gathered by space missions and ground experiments are increasingly used to create training data sets for machine learning algorithms.
These tools are exploited to reach different objectives, as pattern recognition and variable prediction.
Recent examples are the automatic detection of interplanetary coronal mass ejections~[1], the prediction of the North-South component of the magnetic field embedded within interplanetary coronal mass ejections~[2] and the prediction of the global solar radiation~[3].
A powerful exploitation of machine learning models is also the prediction of data gathered by space missions no more active, by using as training set the data taken by an experiment during its lifetime.
An example may be the galactic cosmic-ray (GCR) data gathered on board the European Space Agency LISA Pathfinder (LPF) mission, ended in 2017.
The large amount of available data make it possible to train machine learning models to reproduce the observed GCR flux variations on the basis of contemporaneous and preceding observations of the interplanetary medium, magnetic field and plasma parameters.
Resulting models may be even more valuable if some kind of human-explainable knowledge is provided together with the output predictions.
This goal may be achieved by means of symbolic knowledge-extraction (SKE) techniques, explicitly designed to explain the behaviour of machine learning models in human-interpretable formats.

In this paper we report the results of two SKE techniques, namely \cart{} and \gridex{}, applied to an ensemble predictor reproducing the GCR data gathered on board LPF.
We focus in particular on the readability/fidelity trade-off, to show that it is possible to obtain high degrees of human-intelligibility for the predictions, but at the expense of the corresponding predictive performance.
Accordingly, in \Cref{sec:lpf,sec:ske} some background notions about the LPF mission and SKE techniques, respectively, are provided.
Details about the ensemble predictor and the application of \cart{} and \gridex{} to it are provided in \Cref{sec:contrib}.
Conclusions and possible improvements are presented in \Cref{sec:conclusions}.

\section{The LISA Pathfinder Mission}\label{sec:lpf}

LISA Pathfinder~[4] was an European Space Agency mission aimed at testing if the current technology for the detection in space of gravitational waves with interferometers was mature.
The mission achieved exceptional results, demonstrating the feasibility of placing 2 free falling masses in space with a residual acceleration smaller than a millionth of billionth of the gravitational acceleration.
LPF was the precursor of the scientific mission LISA~[5], which goal will be the detection of supermassive black-hole coalescence.
The LISA mission is scheduled to be launched in 2037.

LPF was launched at the end of 2015 from Kourou (French Guyana) on board a Vega rocket.
Its final orbit around the Lagrangian point L1 was reached on January, 2016 and the mission lifetime ended on July, 2017.
We recall that the L1 point is at 1.5 million km from Earth in the Earth-Sun direction.
Mission orbit was inclined of $\sim45$ degrees w.r.t.\ the ecliptic plane and required approximately six months to be completed by the satellite.
Minor and major axes of the LPF orbit were about 0.5 and 0.8 millions of km, respectively.
The spacecraft rotated around its own axis with a period of six months.

LPF was equipped with a particle detector to monitor the flux of GCRs and particles originated from the Sun energetic enough to traverse the spacecraft and reach the test masses.
The test masses were cubes of platinum and gold, penetrated and charged by protons and ions having energies $>100$ MeV n$^{-1}$.
The test-mass charging induced spurious forces on the test masses~[6].
Monte Carlo simulations have been exploited to study this process before the
mission launch~[7--9].
Noise control has been periodically performed by using ultraviolet light beams to discharge the test masses~[10].
The LPF particle detector enabled the observation of the GCR integral flux with a nominal statistical uncertainty of 1\% on hourly binned data.

\subsection[]{GCR flux short-term variations observed with LISA Pathfinder}

GCR flux short-term variations are characterised by a duration shorter than the solar rotation period ($\sim27$ days) and are associated with the passage of magnetic structures having solar or interplanetary origin.
GCR flux is generally anti-correlated with increasing solar wind speed and interplanetary magnetic field amplitude.
As a consequence, GCR flux, interplanetary magnetic field, solar wind plasma and geomagnetic activity indices show nominal quasi-periodicities related to the Sun rotation period and higher harmonics equal to 27, 13.5 and 9 days~[11, 12].
A quantitative assessment of the correlation between GCR flux depressions and increases of the solar wind speed and/or interplanetary magnetic field intensity is still missing, since the evolution of single short-term depressions is unique and may differ even in presence of similar interplanetary medium conditions.

The particle detector hosted by LPF allowed to study GCR short-term flux variations during Bartels rotations from 2490 through 2509 (from February 18th, 2016 through July 18th, 2017).
We recall that Bartels rotations are 27-day periods defined as complete apparent rotations of the Sun viewed from Earth.
Day 1 of rotation 1 is arbitrarily fixed to February 8, 1832.

GCR percent variations are compared to interplanetary magnetic field intensity and solar wind plasma contemporaneous observations for each Bartels rotations in order to focus on recurrent periodicities consistent with the Sun rotation period and higher harmonics.

\section{Symbolic Knowledge Extraction}\label{sec:ske}

Machine learning models are currently adopted to face a wide variety of tasks, since they exhibit an impressive predictive performance~[13].
These models usually require a prior training phase, during which models learn from training data some kind of generalised knowledge to be used to draw predictions.
A large subset of machine learning predictors store the acquired knowledge in the form of internal parameters (i.e., sub-symbolically), making it difficult for human users to understand the process leading to the model predictions.
These predictors are commonly defined as \emph{opaque}, or \emph{black boxes}~[14].

There exist critical applications (e.g., those having great impact on human lives) that may benefit from the adoption of decision support systems, however these contexts require human awareness about the system internal behaviour.
For this reason systems based on opaque models are not reliable, even if they provide accurate suggestions/predictions.
Amongst the various solutions~[15, 16], the explainable artificial intelligence community proposes SKE methods aimed at explaining the internal functioning and/or the outputs of opaque models~[17].
Amongst the available techniques to achieve the goal of explainability there is the creation of a \emph{surrogate} model, that is a non-opaque predictor able to mimic the opaque one.
In this case, the opaque predictor is called \emph{underlying model}.
Critical applications that benefit from SKE are, for instance, medical diagnosis~[18, 19], credit-risk evaluation~[20--22] and credit card screening~[23].

\cart{}~[24] and \gridex{}~[25] are examples of algorithms applicable to black-box models.
The former induces a decision tree having conditions on the input variables as nodes and output predictions as leaves.
Paths from the tree root to single leaves represent human-intelligible rules mimicking the underlying model predictions.
On the other hand, \gridex{} operates a hyper-cubic partitioning of the input feature space in order to find subregions of instances whose associated output predictions are similar.
Each hyper-cubic region is then described in terms of the input variables (i.e., hyper-cube sides are equivalent to variables whose values lay in specific intervals) and is associated to an output value.
The output value for a hyper-cube is calculated by averaging the predictions provided by the underlying model for the training data belonging to the hyper-cube.
Thus, descriptions are human-understandable and may be used to draw interpretable predictions.

In this work we rely on the \cart{} and \gridex{} implementations included in the \psyke{} framework~[26--28].
\psyke{} is a Python library providing several interchangeable SKE algorithms.
Explainability in \psyke{} is achieved via the extraction of rules in Prolog syntax.

\section{Explaining an Ensemble Model for the LPF GCR Data}\label{sec:contrib}

Since the performance of knowledge extractors is usually bounded to the specific task at hand and there is not a universally best choice, a comparison between the \cart{} and \gridex{} has been performed and reported here.
Both algorithms have been applied to the same ensemble model reproducing the LPF GCR data.
Information about the training data set and the ensemble predictor are also reported in the following.
Design choices about the data set creation and the predictor hyper-parameters have been optimised and described in another work currently not yet published.

\subsection[]{Data set}

A data set has been created to train the ensemble predictor and the extractors.
Since GCR flux variations show a strong correlation with solar wind speed and interplanetary magnetic field intensity, these parameters have been used as data set input variables.
In particular, GCR observations have been temporally aligned with those of magnetic field and solar wind, then for each time instant the following input variables have been selected:
\begin{inlinelist}
	\item 2 variables representing the solar wind speed and the interplanetary magnetic field intensity at the considered time instant;
	\item 6 variables obtained by averaging every 36 hours solar wind speed observations during the 9 days preceding the considered time instant;
	\item 9 variables obtained by averaging every 24 hours interplanetary magnetic field intensity observations during the preceding 9 days;
	\item the GCR flux variation observed 9 days before the considered instant by LPF w.r.t.\ the average monthly value, for GCR normalization.
\end{inlinelist}
The data set has been completed by adding the output variable, that is the GCR flux decrease observed by LPF at the considered instant w.r.t.\ the flux value 9 days before.
As for the number of instances, the data set encompasses all the available observations between the starting and ending time of LPF, resulting in about 11\,000 instances.

\subsection[]{Ensemble model reproducing LPF GCR data}

\begin{figure}[tbh]
	\centering
	\includegraphics[width=.9\linewidth]{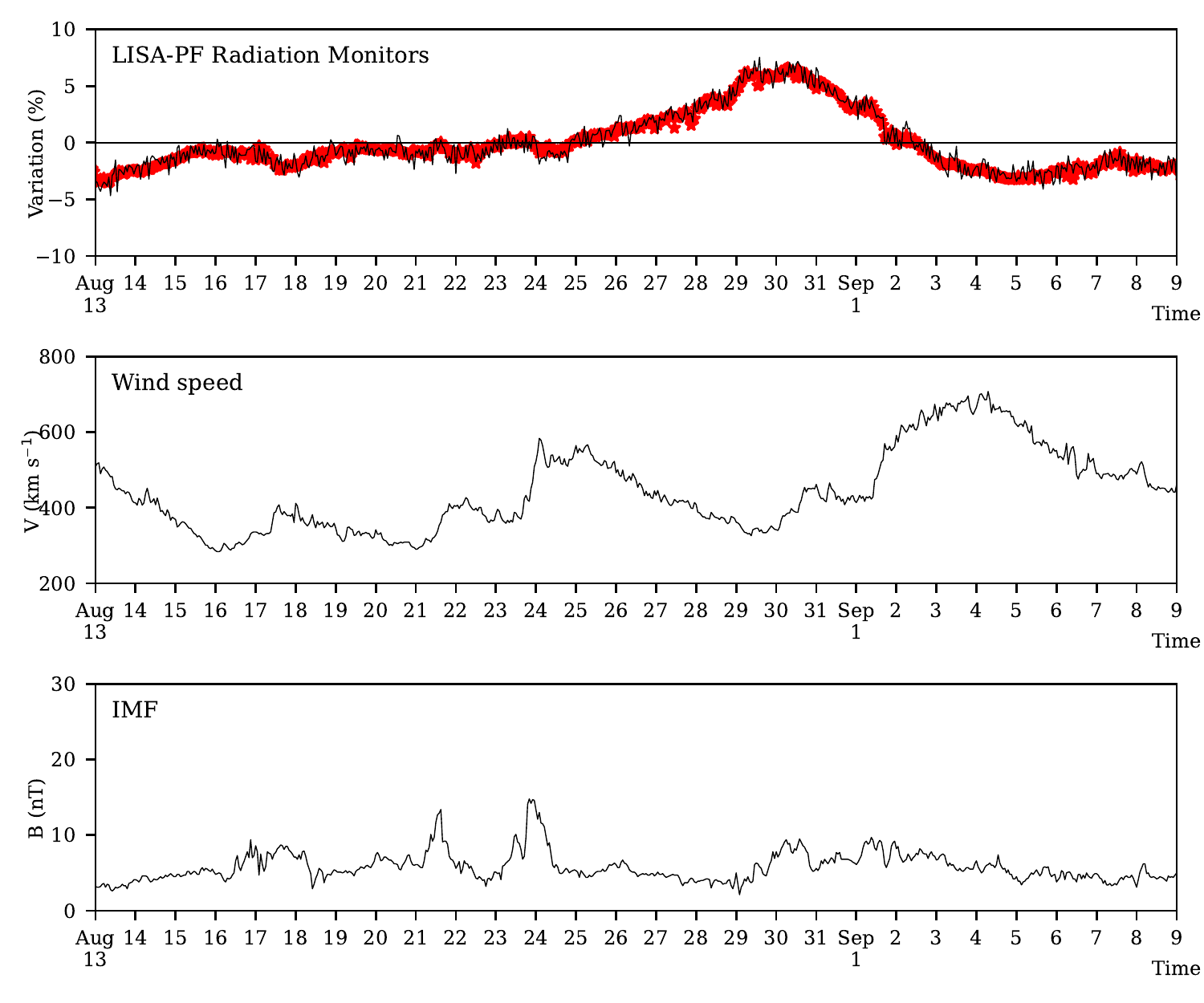}
	\caption{LPF GCR flux data (top panel), solar wind speed (middle panel) and interplanetary magnetic field intensity (bottom panel) observed in L1 during the Bartels rotation 2497. Ensemble model predictions (red stars in the top panel) are reported superposed to the contemporaneous GCR data.}
	\label{fig:example}
\end{figure}

The ensemble model~[29, 30] adopted to reproduce the GCR data observed on board LPF consists of 10 support-vector machines~[31, 32].
These base regressors have been aggregated to obtain a bagging regressor providing output predictions by averaging the base model outputs.
An ensemble approach has been preferred since the exploitation of multiple learning algorithms enables more robust and accurate predictions than single machine learning models.
In this particular case, support-vector machines are supervised learning algorithms requiring 3 hyper-parameters to be tuned:
\begin{inlinelist}
	\item a kernel function,
	\item a regularisation parameter,
	\item a maximum error $\varepsilon$.
\end{inlinelist}
For this work radial basis function kernels have been chosen as kernel functions, i.e., non-linear generalised kernels commonly exploited to perform linear separations of non-linearly separable data.
This is achieved by mapping the input features into a new input space having higher dimensionality.
The regularisation parameter, controlling the tolerance for outliers, has been fixed to 1.
Finally, the maximum error $\varepsilon$ has been chosen equal to 0.1, to not penalise predictions having error $<0.1$ during the training phase of the models.

Since our goal is the knowledge extraction from the ensemble model, and not to obtain future predictions, all the available data have been used to train the model, without keeping apart a test set.
This allowed us to obtain a predictor perfectly reproducing the GCR data gathered by LPF.
\Cref{tab:predictor} reports the mean absolute error measured for the ensemble model outputs averaged per Bartels rotation.
In the Table are reported all the Bartels rotation during which LPF gathered data.
As an example, model outputs are also reported in \Cref{fig:example} for the Bartels rotation 2497.

\setlength{\tabcolsep}{0.5em}
\begin{table}[tbp]
	\caption{Mean absolute error (MAE) and standard deviation measured for the ensemble model outputs for each Bartels rotation during the LPF mission.}
	\begin{center}
		\begin{tabular}{cc|cc}
			\toprule
			Bartels rotation & MAE (\%) & Bartels rotation & MAE (\%) \\
			\midrule
			2491 & 0.52 $\pm$ 0.42 & 2500 & 0.48 $\pm$ 0.38 \\
			2492 & 0.53 $\pm$ 0.41 & 2501 & 0.49 $\pm$ 0.37 \\
			2493 & 0.54 $\pm$ 0.43 & 2502 & 0.47 $\pm$ 0.35 \\
			2494 & 0.50 $\pm$ 0.40 & 2503 & 0.46 $\pm$ 0.36 \\
			2495 & 0.51 $\pm$ 0.40 & 2504 & 0.45 $\pm$ 0.34 \\
			2496 & 0.49 $\pm$ 0.38 & 2505 & 0.43 $\pm$ 0.31 \\
			2497 & 0.50 $\pm$ 0.38 & 2506 & 0.51 $\pm$ 0.42 \\
			2498 & 0.50 $\pm$ 0.38 & 2507 & 0.49 $\pm$ 0.39 \\
			2499 & 0.51 $\pm$ 0.38 & 2508 & 0.46 $\pm$ 0.37 \\
			All & \multicolumn{3}{l}{0.49 $\pm$ 0.38} \\
			\bottomrule
		\end{tabular}
	\end{center}
	\label{tab:predictor}
\end{table}

\subsection[]{Knowledge extraction}

The trained ensemble model is used as underlying predictor for the \cart{} and \gridex{} extractors.
Knowledge extraction from the model is preferred over direct rule induction from the data set, since it enables the adoption of the underlying model as an oracle to augment the training set.

\subsubsection[]{CART}

\begin{figure}[tbh]
	\centering
	\includegraphics[width=.9\linewidth]{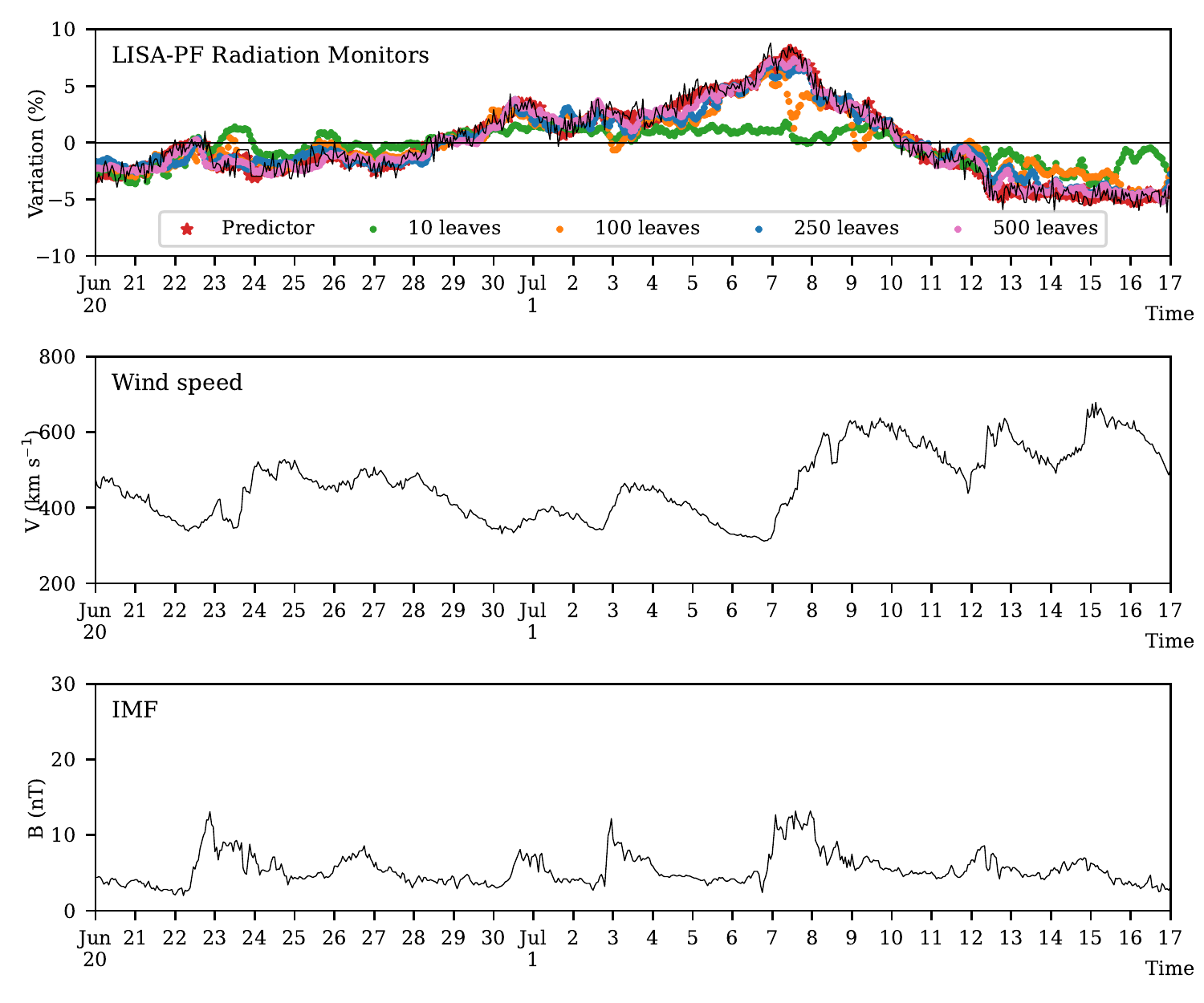}
	\caption{\cart{} output predictions for different values of the maximum leaf amount (equal to 10, 100, 250 and 500) (top panel) for the Bartels rotation 2495. Panels are the same as in \Cref{fig:example}.}
	\label{fig:cart}
\end{figure}

The predictive performance of \cart{} strictly depends on the number of output leaves of the induced decision tree.
Indeed, \cart{} provides constant predictions and introduces an undesired output discretisation.
The effects of the discretisation are limited if the output variable to be predicted is discrete or if there is a large amount of leaves.
However, a large amount of leaves, and thus of output human-readable rules, hinder the readability of the model.

We trained several instances of \cart{}, with different values for the maximum allowed number of leaves.
The readability/fidelity trade-off is problematic in the present applicative context, since it is possible to obtain a good fidelity (small predictive error w.r.t.\ the ensemble model predictions) only with a very large amount of leaves.
In particular, a good agreement between LPF GCR data and \cart{} predictions can be found with more than 200 leaves.
Readability of \cart{} rules is also hindered by the number of antecedents per rule.
Whereas shallow leaves are translated into rules having few antecedents, deeper leaves may result in rules having too many conditions to be still considered as human-readable.
Since the number of antecedents in a rule is equal to the associated leaf depth, it is possible to limit this drawback by setting a maximum depth in the tree induction.
However, this in turn worsens the extracted rule fidelity.

In \Cref{tab:cart} the predictive performance and the number of leaves of the tested \cart{} instances are reported.
Predictive performance is expressed as mean absolute error with respect to both the data and the ensemble model predictions.
A visual comparison of \cart{} with 10, 100, 250 and 500 leaves is reported in \Cref{fig:cart} for the Bartels rotation 2495.

\begin{table}[tbp]
	\caption{Number of leaves and mean absolute error measured for \cart{} w.r.t.\ the data and the ensemble model predictions.}
	\begin{center}
		\begin{tabular}{c|cc}
			\toprule
			\# of leaves & \multicolumn{2}{c}{MAE (\%)} \\
			& Data & Model \\
			\midrule
			10 & 1.70 $\pm$ 1.32 & 1.57 $\pm$ 1.24 \\
			25 & 1.52 $\pm$ 1.20 & 1.39 $\pm$ 1.11 \\
			50 & 1.37 $\pm$ 1.07 & 1.23 $\pm$ 0.98 \\
			100 & 1.18 $\pm$ 0.94 & 1.03 $\pm$ 0.85 \\
			150 & 1.07 $\pm$ 0.85 & 0.91 $\pm$ 0.75 \\
			250 & 0.91 $\pm$ 0.74 & 0.74 $\pm$ 0.62 \\
			500 & 0.73 $\pm$ 0.58 & 0.51 $\pm$ 0.43 \\
			\bottomrule
		\end{tabular}
	\end{center}
	\label{tab:cart}
\end{table}

Examples of \cart{} rules are the following:
\begin{lstlisting}
GCR flux increment is 0.91% if
  GCR0 < 2.52%,    V < 344,    V1 < 429.
	
GCR flux decrement is 4.15% if
  GCR0 < 6.51%,    V  < 621,    V1 < 649,    V3 < 683,
  V5   < 649,      V6 < 690,    B1 < 13.8.
\end{lstlisting}
Variable \verb|GCR0| represents the GCR flux value 9 days before the considered instant.
\verb|V| is the solar wind speed in the same instant, while \verb|Vi| is the average solar wind speed value taken in the \verb|i|-th time window before the considered instant.
Wind speed is always expressed in km s$^{-1}$ and time windows have dimensions equal to 36 hours.
The same holds for \verb|B1|, that represents the interplanetary magnetic field intensity averaged in the 24 hours before the considered instant.
Magnetic field is expressed in nT.
It is evident how the first rule, with only 3 antecedents, is more readable than the second, having 7 antecedents.

\subsubsection[]{GridEx}

\begin{figure}[tbh]
	\centering
	\includegraphics[width=.9\linewidth]{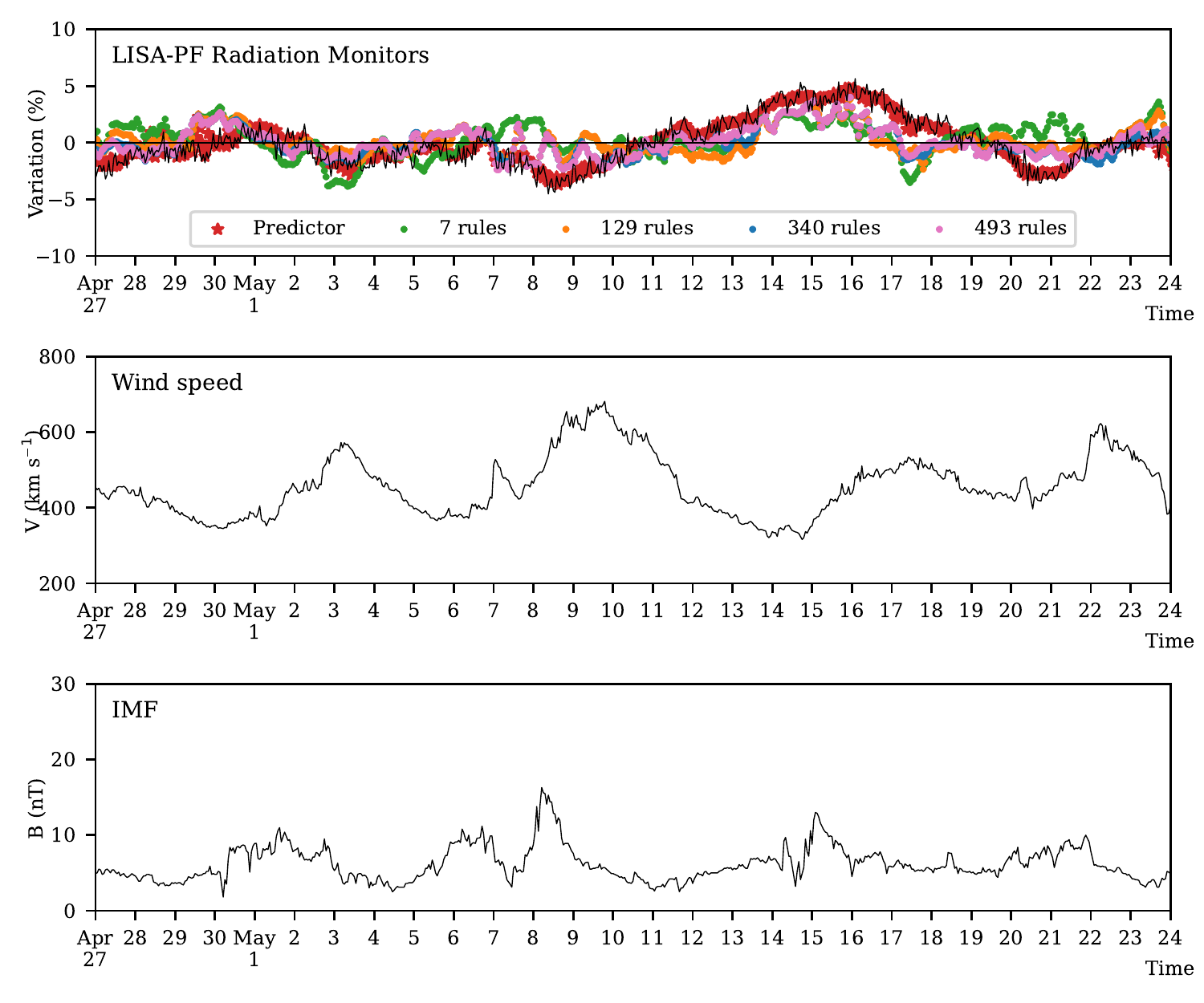}
	\caption{\gridex{} output predictions for different values of its parameters, resulting in different amounts of output rules (top panel) for the Bartels rotation 2493. Panels are the same as in \Cref{fig:example}.}
	\label{fig:gridex}
\end{figure}

To bypass the readability limitations of \cart{} also the \gridex{} algorithm has been applied to the ensemble predictor.
Thanks to the adaptive splitting of \gridex{} it is possible to create output rules by only involving the most relevant input features in the precondition, resulting in a more controlled rule readability.
\gridex{} requires the fine tuning of a set of parameters, namely:
\begin{inlinelist}
	\item the minimum amount of training instances to be considered in each input space region;
	\item the depth of the recursive partitioning;
	\item the error threshold for selecting regions to be further partitioned;
	\item the splitting strategy.
\end{inlinelist}
In all our experiments, the minimum amount of samples has been fixed equal to 100.
This means that if less than 100 training instances are included in an input space subregion, during the training phase of \gridex{} the data set is augmented by generating random input samples inside the region and predicting them by using the underlying ensemble model as an oracle.
The depth of the partitioning controls how many times the input space subregions have to be split.
Regions are split only if the predictive error inside them is greater than the user-defined error threshold.
Finally, adaptive strategies have been chosen as splitting strategies.
In particular, in all the experiments the less relevant input features have not been split, while the more relevant ones have been split into 2 partitions at each iteration.
Features are considered as \emph{more relevant} if their relevance is greater than a relevance threshold, not constant for all the experiments.
The parameter values adopted for the experiments are reported in \Cref{tab:gridex}.
A visual comparison of different instances of \gridex{} is reported in \Cref{fig:gridex} for the Bartels rotation 2493.

\begin{table}[tb]
	\caption{Parameters, number of extracted rules and mean absolute error measured for several instances of \gridex{} w.r.t.\ the data and the ensemble model predictions.}
	\begin{center}
		\begin{tabular}{ccc|ccc}
			\toprule
			Depth & Error & Relevance & \# of rules & \multicolumn{2}{c}{MAE (\%)} \\
			& threshold & threshold & & Data & Model \\
			\midrule
			1 & 0.60 & 0.1 & 7 & 2.05 $\pm$ 1.54 & 1.95 $\pm$ 1.47 \\
			2 & 0.60 & 0.1 & 24 & 1.86 $\pm$ 1.40 & 1.74 $\pm$ 1.33 \\
			3 & 0.60 & 0.1 & 49 & 1.74 $\pm$ 1.32 & 1.62 $\pm$ 1.24 \\
			3 & 0.60 & 0.05 & 129 & 1.70 $\pm$ 1.30 & 1.58 $\pm$ 1.23 \\
			3 & 0.55 & 0.05 & 219 & 1.58 $\pm$ 1.23 & 1.45 $\pm$ 1.15 \\
			3 & 0.50 & 0.05 & 340 & 1.45 $\pm$ 1.16 & 1.31 $\pm$ 1.08 \\
			3 & 0.45 & 0.05 & 414 & 1.40 $\pm$ 1.14 & 1.26 $\pm$ 1.06 \\
			3 & 0.40 & 0.05 & 493 & 1.36 $\pm$ 1.12 & 1.21 $\pm$ 1.04 \\
			\bottomrule
		\end{tabular}
	\end{center}
	\label{tab:gridex}
\end{table}

Examples of \gridex{} output rules are the following.
\begin{lstlisting}
GCR flux increment is 2.75% if
  -7.72 % < GCR0 < 0.73%,   263 < V < 508,   285 < V1 < 502.
	
GCR flux decrement is 5.28% if
  0.73 % < GCR0 < 9.19%,   508 < V < 752,   502 < V1 < 720.
\end{lstlisting}
Variables follow the same convention as for \cart{} output rules.
In this case all the extracted rules have the same readability, since the number of rule antecedents is fixed to 3 via the splitting strategy parameter tuning.

\subsubsection[]{Comparison between CART and GridEx}

By comparing the output models obtained via the \cart{} and \gridex{} extractors, it is possible to notice that
\begin{inlinelist}
	\item \gridex{} rules have a better readability than \cart{} ones in terms of antecedents per rule; conversely,
	\item fidelity and predictive performance of \gridex{} are worse than those of \cart{}.
\end{inlinelist}
Indeed, even by accepting growing amounts of output rules, \gridex{} is not able to provide predictions with sufficient quality.
We believe that better results may be achieved by using an algorithm able to approximate local predictions with non-constant outputs, as for instance a linear combination of the input variables.
At the moment an extraction algorithm capable of doing so when applied to a complex underlying predictor (as an ensemble model) is still missing in the literature.

A comparison of the mean absolute error measured for both \cart{} and \gridex{} w.r.t.\ the data as well as the ensemble model predictions is reported in \Cref{fig:comparison}.
It is clearly noticeable that \cart{} performs better than \gridex{} and that the performance of the latter does not sensibly improve by increasing the number of extracted rules.

\begin{figure}[tbh]\centering
	\subfloat[MAE (data).]{
		\includegraphics[width=0.49\linewidth]{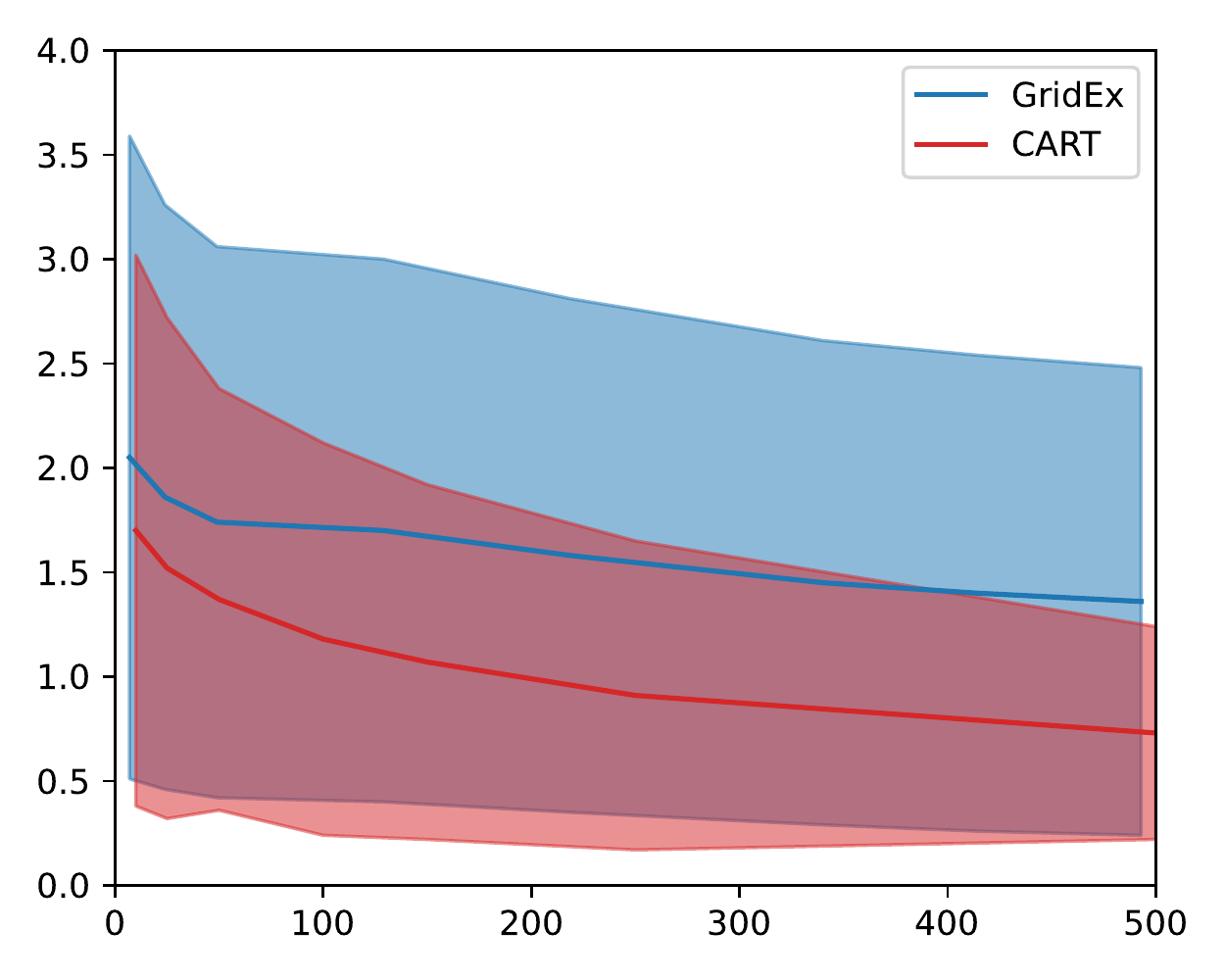}\label{fig:comp1}
	}
	\subfloat[MAE (model).]{
		\includegraphics[width=0.49\linewidth]{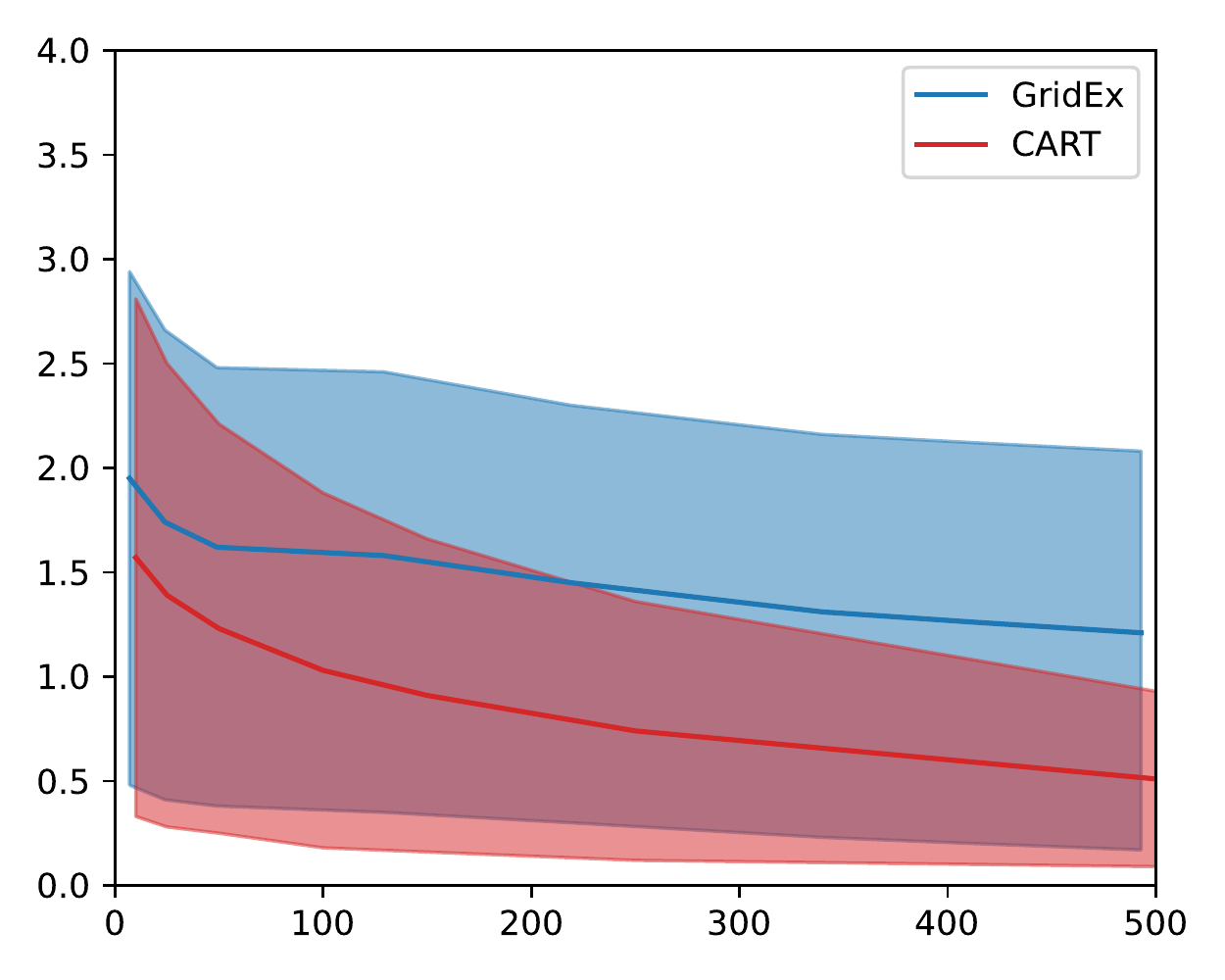}\label{fig:comp2}
	}
	\caption{Comparison between the predictive performance w.r.t.\ the data and the fidelity w.r.t.\ the ensemble model measured via mean absolute error for \cart{} and \gridex{}.}\label{fig:comparison}
\end{figure}

\section{Conclusions}\label{sec:conclusions}

In this work SKE techniques have been applied to a machine learning ensemble model capable of reproducing the GCR data gathered by LPF with an error smaller than the nominal uncertainty of LPF GCR hourly binned data.
Namely, \cart{} and \gridex{} have been applied to the model in order to extract human-readable rules expressing the intensity of GCR flux variations.
The adopted extractors are able to provide good predictions only in limited regions of the input feature space.
This is reasonable since the peculiarities of the interplanetary structures and solar wind high-speed streams make impractical a global approximation of the GCR flux variations for all the possible solar wind speed and interplanetary magnetic field intensity input values~[33].
Thus, only local human-readable approximations can be suited to substitute opaque predictions of an ensemble model.
In our future works we plan to enhance the extraction of knowledge from machine learning models reproducing the LPF GCR data by obtaining fewer rules with smaller predictive error, i.e., we plan to obtain higher degrees of readability and fidelity.
This goal can be achieved by substituting the constant output values of the extracted rules with local linear combinations (or other kinds of functions) of the input variables.
}

\section*{References}

\newcommand{\journal}[5]{\item #1. #2. \emph{#3}. #4. \url{#5}.}

\newcommand{\issue}[8]{\item #1. #2. \emph{#3}. #4;#5(#6):#7. \url{#8}.}

\newcommand{\issuenop}[7]{\item #1. #2. \emph{#3}. #4;#5(#6). \url{#7}.}

\newcommand{\noissue}[7]{\item #1. #2. \emph{#3}. #4;#5:#6. \url{#7}.}

\newcommand{\proc}[9]{\item #1. #2. In: #3, editors. #4; #5; #6. #7. p. #8. \url{#9}.}

\newcommand{\book}[7]{\item #1. \emph{#2}. #3 ed. #4: #5; #6. \url{#7}.}

\newcommand{\booknoed}[6]{\item #1. \emph{#2}. #3: #4; #5. \url{#6}.}

\newcommand{\bookchapter}[8]{\item #1. #2. In: \emph{#3}. #4: #5; #6. p. #7. \url{#8}.}

{
\mysize{8pt}{7pt}

\begin{enumerate}[label={[\arabic*]}]
	\journal{R{\"u}disser HT et al}{Automatic detection of interplanetary coronal mass ejections in solar wind in situ data}{arXiv preprint arXiv:2205.03578}{2022}{https://arxiv.org/abs/2205.03578}
	
	\issue{Reiss MA et al}{Machine learning for predicting the B$_\text{z}$ magnetic field component from upstream in situ observations of solar coronal mass ejections}{Space Weather}{2021}{19}{12}{e2021SW002859}{ https://doi.org/10.1029/2021SW002859}
	
	\noissue{Zhou Y et al}{A review on global solar radiation prediction with machine learning models in a comprehensive perspective}{Energy Convers. Manag}{2021}{235}{113960}{https://doi.org/10.1016/j.enconman.2021.113960}
	
	\issue{Armano M et al}{{LISA Pathfinder}: The experiment and the route to {LISA}}{Class. Quantum Gravity}{2009}{26}{9}{094001}{https://doi.org/10.1088/0264-9381/26/9/094001}
	
	\journal{Amaro-Seoane P et al}{Laser interferometer Space Antenna}{arXiv preprint arXiv:1702.00786}{2017}{https://arxiv.org/abs/1702.00786}
	
	\proc{Shaul DNA et al}{Solar and cosmic ray physics and the space environment: Studies for and with LISA}{Merkowitz SM, Livas JC}{American Institute of Physics. AIP 2006: Proceedings of the 6th International LISA Symposium}{2006 June 19--23; Greenbelt, Maryland (USA)}{United States: AIP}{2006}{172--178}{http://aip.scitation.org/doi/abs/10.1063/1.2405038}
	
	\issue{Ara{\'{u}}jo HM et al}{Detailed calculation of test-mass charging in the LISA mission}{Astropart. Phys.}{2005}{22}{5-6}{451--469}{https://doi.org/10.1016/j.astropartphys.2004.09.004}
	
	\issue{Grimani C et al}{LISA test-mass charging process due to cosmic-ray nuclei and electrons}{Class. Quantum Gravity}{2005}{22}{10}{S327--S332}{https://iopscience.iop.org/article/10.1088/0264-9381/22/10/025}
	
	\issue{Grimani C et al}{LISA Pathfinder test-mass charging during galactic cosmic-ray flux short-term variations}{Class. Quantum Gravity}{2015}{32}{3}{035001}{https://iopscience.iop.org/article/10.1088/0264-9381/32/3/035001}
	
	\issue{Armano M et al}{Charge-induced force noise on free-falling test masses: Results from LISA Pathfinder}{Phys. Rev. Lett.}{2017}{118}{17}{171101}{http://link.aps.org/doi/10.1103/PhysRevLett.118.171101}
	
	\issue{Storini M, Iucci N, Pase S}{North-south anisotropy during the quasi-stationary modulation of galactic cosmic rays}{Il Nuovo Cimento C}{1992}{15}{5}{527--538}{https://link.springer.com/article/10.1007/BF02507827}

	\issuenop{Sabbah I, Kudela, K}{Third harmonic of the 27 day periodicity of galactic cosmic rays: Coupling with interplanetary parameters}{J. Geophys. Res. Space Phys.}{2011}{116}{A4}{https://agupubs.onlinelibrary.wiley.com/doi/full/10.1029/2010JA015922}
	
	\noissue{Rocha A, Papa JP, Meira LAA}{How far do we get using machine learning black-boxes?}{	Int. J. Pattern Recognit. Artif. Intell.}{2012}{26}{1--23}{https://doi.org/10.1142/S0218001412610010}
	
	\issue{Lipton ZC}{The mythos of model interpretability}{Queue}{2018}{16}{3}{31--57}{https://doi.org/10.1145/3236386.3241340}
	
	\issue{Guidotti R et al}{A survey of methods for explaining black box models}{ACM Comput. Surv.}{2018}{51}{5}{1--42}{https://doi.org/10.1145/3236009}
	
	\issue{Rudin C}{Stop explaining black box machine learning models for high stakes decisions and use interpretable models instead}{Nat. Mach. Intell.}{2019}{1}{5}{206--215}{https://doi.org/10.1038/s42256-019-0048-x}
	
	\noissue{Kenny EM et al}{Explaining black-box classifiers using post-hoc explanations-by-example: The effect of explanations and error-rates in {XAI} user studies}{Artif. Intell.}{2021}{294}{103459}{https://doi.org/10.1016/j.artint.2021.103459}

	\issue{Hayashi Y, Setiono R, Yoshida K}{A comparison between two neural network rule extraction techniques for the diagnosis of hepatobiliary disorders}{Artif. Intell. Med.}{2000}{20}{3}{205--216}{https://doi.org/10.1016/s0933-3657(00)00064-6}

	\noissue{Bologna G, Pellegrini C}{Three medical examples in neural network rule extraction}{Phys. Med.}{1997}{13}{183--187}{https://archive-ouverte.unige.ch/unige:121360}

	\proc{Baesens B et al}{Building credit-risk evaluation expert systems using neural network rule extraction and decision tables}{Storey VC, Sarkar S, DeGross JI}{International Conference in Information Systems. ICIS 2001: ICIS 2001 Proceedings}{2001 June 18--19; New Orleans, LA (USA)}{United States: Association for Information Systems}{2001}{159--168}{http://aisel.aisnet.org/icis2001/20}

	\issue{Baesens B et al}{Using neural network rule extraction and decision tables for credit-risk evaluation}{Manag. Sci.}{2003}{49}{3}{312--329}{https://doi.org/10.1287/mnsc.49.3.312.12739}

	\issue{Steiner MTA et al}{Using neural network rule extraction for credit-risk evaluation}{Int. J. Netw. Secur.}{2006}{6}{5A}{6--16}{https://citeseerx.ist.psu.edu/viewdoc/download?doi=10.1.1.570.4480&rep=rep1&type=pdf}
	
	\issue{Setiono R, Baesens B, Mues C}{Rule extraction from minimal neural networks for credit card screening}{Int. J. Neural Syst.}{2011}{21}{04}{265--276}{https://doi.org/10.1142/S0129065711002821}
	
	\book{Breiman L et al}{Classification and regression trees}{1st}{New York}{Routledge}{1984}{https://www.routledge.com/Classification-and-Regression-Trees/Breiman-Friedman-Stone-Olshen/p/book/9780412048418}
	
	\proc{Sabbatini F, Ciatto G, Omicini A}{{GridEx}: An algorithm for knowledge extraction from black-box regressors}{Calvaresi D, Najjar A, Winikoff M, Fr{\"a}mling K}{EXTRAAMAS 2021: Proceedings of the 3rd International Workshop on EXplainable and TRAnsparent AI and Multi-Agent Systems}{2021 May 3--7; London (United Kingdom}{Cham: Springer}{2021}{18--38}{https://link.springer.com/chapter/10.1007/978-3-030-82017-6_2}
	
	\proc{Sabbatini F et al}{On the design of {PSyKE}: A platform for symbolic knowledge extraction}{Calegari R, Ciatto G, Denti E, Omicini A, Sartor G}{WOA 2021: Proceedings of the 22nd Workshop ``From Objects to Agents''}{2021 September 1--3; Bologna (Italy)}{Aachen: Sun SITE Central Europe, RWTH Aachen University}{2021}{29--48}{http://woa2021.apice.unibo.it/paper/WOA_2021_paper_14.pdf}
	
	\issue{Sabbatini F et al}{Symbolic knowledge extraction from opaque {ML} predictors in {PSyKE}: Platform design \& experiments}{Intelligenza Artificiale}{2022}{16}{1}{27--48}{https://content.iospress.com/articles/intelligenza-artificiale/ia210120}
	
	\item Sabbatini F, Ciatto G, Omicini A. {Semantic Web}-based interoperability for intelligent agents with {PSyKE}. EXTRAAMAS 2022: Proceedings of the 4th International Workshop on EXplainable and TRAnsparent AI and Multi-Agent Systems; 2022 May 9--10; Auckland, NZ (fully online), in press.
	
	\proc{Dietterich TG}{Ensemble Methods in Machine Learning}{Kittler J, Roli F}{Multiple Classifier Systems. MCS 2000: Proceedings of the First International Workshop on Multiple Classifier Systems}{2000 June 21--23; Cagliari (Italy)}{Berlin: Springer}{2000}{1--15}{https://link.springer.com/book/10.1007/3-540-45014-9}
	
	\book{Zhou Z-H}{Ensemble methods: Foundations and algorithms}{1st}{United Kingdom}{Chapman and Hall/CRC}{2012}{https://dl.acm.org/doi/10.5555/2381019}
	
	\bookchapter{Suthaharan S}{Support vector machine}{Machine learning models and algorithms for big data classification: Thinking with examples for effective learning}{New York}{Springer}{2016}{207--235}{https://link.springer.com/book/10.1007/978-1-4899-7641-3}
	
	\booknoed{Steinwart I, Christmann A}{Support vector machines}{Berlin}{Springer}{2008}{https://link.springer.com/book/10.1007/978-0-387-77242-4}
	
	\issuenop{Grimani C et al}{Recurrent galactic cosmic-ray flux modulation in {L1} and geomagnetic activity during the declining phase of the solar cycle 24}{Astrophys. J.}{2020}{904}{1}{https://doi.org/10.3847/1538-4357/abbb90}

\end{enumerate}
}

%\bibliographystyle{apalike}
%\bibliography{AAOAJ-2022-LPFSKE}

%\printbibliography

\end{document}